\documentclass[prc,unsortedaddress,superscriptaddress,twocolumn,amsmath,amsfonts,amssymb,showpacs,floatfix]{revtex4}
\usepackage{epsfig}

\begin{document}

\title{Bound, virtual and resonance $S$-matrix poles from the Schr\"odinger equation}

\author{A.\,M.~Mukhamedzhanov}
\affiliation{Cyclotron Institute, Texas A\&M University, College
Station, TX 77843}

\author{B.\,F.~Irgaziev}
\affiliation{GIK Institute of Engineering Sciences and Technology,
Topi, Pakistan}

\author{V.\,Z.~Goldberg}
\affiliation{Cyclotron Institute, Texas A\&M University, College
Station, TX 77843}

\author{Yu.\,V.~Orlov}
\affiliation{Institute of Nuclear Physics, Moscow State
University, Moscow, Russia}

\author{I.\,~Qazi }
\affiliation{GIK Institute of Engineering Sciences and Technology,
Topi, Pakistan}

\date{\today}

\begin{abstract}
A general method, which we call the potential $S$-matrix pole method, is
developed for obtaining the $S$-matrix pole
parameters for bound, virtual and resonant states based on
numerical solutions  of the Schr\"odinger equation.
This method is well-known for bound states. In this
work we generalize it for resonant and virtual states,
although the corresponding solutions increase
exponentially when $r\to\infty$. Concrete calculations are
performed for the $1^+$ ground  and the $0^+$ first excited states
of $^{14}\rm{N}$,  the resonance $^{15}\rm{F}$ states ($1/2^+$,
$5/2^+$), low-lying states of $^{11}\rm{Be}$ and $^{11}\rm{N}$,
and the subthreshold resonances in the proton-proton system.
We also demonstrate that in the case the broad
resonances their energy and width can be found from the fitting of the  experimental phase shifts using the analytical expression for the elastic scattering $S$-matrix.  We compare
the $S$-matrix pole and the $R$-matrix for broad $s_{1/2}$ resonance in ${}^{15}{\rm F}$.
\end{abstract}

\maketitle
\section{Introduction}
Analysis of the $S$-matrix pole structure is a powerful method in
quantum physics. It is well-known that the poles of the $S$-matrix in
the complex momentum (or energy) plane correspond to bound,
virtual and resonance states. There is a well-known relation
between the $S$-matrix and Jost functions, the singular
solutions of the Schr\"odinger equation at $r\to 0$. The
conventional numerical method for bound states is to search for
solutions, which only have an outgoing wave at pure imaginary
momenta in the upper half momentum plane. The corresponding wave
function is an exponentially decreasing solution when $r\to
\infty$.

Virtual or resonance states are described by the wave functions
containing only the outgoing waves asymptotically, which
exponentially increase due to the complex momenta. In the
past (see \cite{orl89} and references therein), the analytical
continuation onto the unphysical energy sheet of the
Lippmann--Schwinger as well as the momentum-space Faddeev integral
equations were used to find the resonance properties. The
normalization formula for the bound state vertex function in the
momentum space was generalized  in \cite{orl84} for the
resonance and virtual states.

Such states are considered as unaccomplished bound states.
This means that a bound state pole should move down the positive
semi-axis of the complex momentum plane while the interaction
strength decreases. At some critical value of the interaction
strength, the pole approaches the zero energy point, which belongs
to the  contour of integration. After a subsequent decrease of the
interaction strength, the pole moves to the lower part of the
complex momentum plane (unphysical energy sheet) dragging the
integration contour in the Lippmann--Schwinger equation to secure
the convergence of the integral. This leads to the appearance of
an extra term in the right-hand-side of the equation, which is the
residue of the integrand at the pole.

This method of the analytical continuation has been applied
successfully to different physical systems. Unfortunately, it can not be
used directly in the case of charged particles. We should also
note that an analytical form of the Fourier transform of the potential,
which is an input in the Lippmann-Schwinger integral equations, is known
only for a limited number of potentials.

The problem of the exponential increase of the Gamow resonance
wave function in the asymptotic region can be solved by a complex
scaling method based on the so-called ABC$-$theorem \cite{agu71}.
This method consists of solving the Schr\"odinger equation on a
ray in the first quadrant of the radial complex plane rather than
on the real axis of the coordinate \textit{r}. This ray can be
obtained by the following transformation of the radial coordinate
\textit{r} and the conjugate momentum \textit{p}: $r\to
r\,\exp(i\theta ) $ and $p\to p\,\exp(-i\theta)$. As a result, the
bound state spectrum is supplemented by the $S$-matrix poles
situated in the sector defined by the angle $\theta$ between the
real axis and the ray in the fourth quadrant of the complex
momentum plane. The axis rotation angle, $\theta$, is limited by the
position of the potential singularities in the radial complex
plane. It is important that the complex scaling method can be
applied to the case of charged particles. The method is valid
because the Coulomb potential satisfies the scaling condition of
the ABC$-$theorem. An application of this method to resonances in
nuclear reactions was presented in \cite{gua86}. The numerical
realization of this method is a rather complex one.

A few different techniques to determine the resonance energy, width
and resonance wave function based on the solution of the Schr\"odinger
equation have been previously suggested. In these methods the normalization of the resonant wave function is achieved  using the Zel'dovich's
normalization \cite{zel60}, which is difficult in practical realization due to slow convergence of the integrals.  First we refer to the method of solution of the radial Schro\"odinger equation to determine resonances
suggested in \cite{vertse82}. In this method the complex eigenvalue and the Gamow wave function can be found by integration of the Schr\"odinger
equation imposing the boundary conditions in the origin and the
asymptotic region. To solve the equation the Fox-Goodwin numerical
method was applied and the logarithmic derivatives of the
internal and external wave functions were matched. However, this method
fails in the vicinity of the threshold, for broad and subthreshold broad resonances (imaginary part
of the momentum is larger than its real part) and antibound (virtual)
states. We underscore also that this method can be applied only for
the potentials with the finite interaction radius because of the
problem with numerical calculation of the exponentially increasing
wave function. The application of the method \cite{vertse82} for the unstable
nuclei can be found in \cite{vertse88,mil2001}.

A pole search has also been used in \cite{nazar2003} by solution of the Schr\"odinger equation
with the short range interaction for the scattering wave function. The Zel'dovich's normalization  procedure for the Gamow resonance wave function supplemented by the exterior complex scaling\cite{gyarmativertse} was used. The norm of the Gamow resonant wave functions
does exists for charged particles also \cite{gyarmativertse,dol77}.
The method allows one to find resonances and even subthreshold resonances but it cannot 
be applied to the virtual states.

The method, which is also close to our approach, was discussed in
\cite{csoto98}. The asymptotic wave function in this method
contains auxiliary $\tilde{S}$-matrix which coincides with the
physical $S$-matrix at the resonance poles at which the solution
becomes pure outgoing wave. The method was applied for
determination of the low-energy $^5\rm{He}$ and $^5\rm{Li}$
resonance parameters \cite{csoto98}.

In the present work, we demonstrate how to find the poles of the
$S$-matrix corresponding to bound, virtual and resonance
states and the $S$-matrix residues in these poles by solving the Schr\"odinger
equation with the nuclear plus Coulomb potentials using the analytical properties of the $S$-matrix.
In contrast to the previously published methods, in our $S$-matrix pole method
the normalization of the resonant wave function is based on the connection
between the residue of the $S$-matrix in the pole and the asymptotic normalization coefficient (ANC). This relationship is universal and can be applied to bound, virtual and resonance (narrow and broad) states \cite{dol77} making our technique universal, and that is the main distinction of our method from the previously published ones. The ANC is the amplitude of the tail of the bound, virtual or resonant wave function \cite{dol77,blokh77}. For the resonant state, the ANC is
related to the resonance width \cite{dol77}. The use of the ANC doesn't require
the normalization of the state corresponding to the $S$-matrix pole and this why our method allows one to determine both narrow and broad resonances, and even antibound states.

 A simple relation between the ANC (nuclear vertex constant) and the overall normalization of the peripheral astrophysical $S$-factor  suggested in \cite{muk90,xu94,mukh01} makes it extremely important for obtaining astrophysical $S$ factors. Note that the normalization method proposed by Zel'dovich \cite{zel60} was generalized in \cite{dol77} for the interaction potential with a Coulomb tail.

The $S$-matrix pole method addressed here has been applied earlier to the
virtual (singlet) deuteron and virtual triton with different
short-range potentials. The results of the two-step Gamov state 
normalization for  the virtual (antibound) state of ${}^3\rm{H}$ were 
published in \cite{irg06}. For charged particles, the virtual state 
becomes a subthreshold resonance \cite{kok80}. Here we present new results for the subthreshold
resonances for the ground state of ${}^{2}{\rm H}$. We also
calculate the ground and the first excited states of
${}^{14}\rm{N}$ and the resonance states of ${}^{15}\rm{F}$.
Finally, our method is applied to the three lowest $T=\frac{3}{2}$
states in ${}^{11}\rm{Be}$ and ${}^{11}\rm{N}$. Considering
the $\frac{1}{2}^{+}$ state in $^{11}\rm{N}$ as an example, we demonstrate
how to determine in a model-independent way the energy and width of the broad resonance using the $S$-matrix analytical structure, which includes the resonant pole. Moreover, we demonstrate that the potential $S$-matrix pole method addressed
here gives the resonance energy and width, which are very close to the model-independent results obtained from the analytical expression for the $S$-matrix in the vicinity of a single pole \cite{baz}.

We use the system of units in which $\hbar$=$c$=1.

\section{A NUMERICAL CALCULATION METHOD}

To describe the nuclear interaction we adopt the Woods-Saxon
potential
\begin{equation}
V_{N} (r)=-[V_{0} -V_{LS} (\vec{L}\cdot \vec{S})\frac{2}{m_{\pi
}^{2} } \frac{d}{rdr} ]\frac{1}{1+\exp [\frac{r-R_{N}}{a}]} ,
\label{GrindEQ__1_}
\end{equation}
where $V_0$ ($V_{LS }$) is the depth of the central (spin-orbital)
potential; $\vec{L}$ is the orbital momentum operator for the
relative motion of the particles; $\vec{S}$ is the spin operator;
$m_\pi$ is the pion mass; \textit{a} is the diffuseness and
$R_N=r_0A^{1/3}$ ($r_0$ is the radius parameter of the nuclear
potential, \textit{A} is the atomic mass number). The Coulomb
interaction potential is taken in the form
\begin{equation} \label{GrindEQ__2_}
V_{C} (r)=\left\{\begin{array}{c} {\frac{Z_{1} Z_{2} e^{2} }
{2R_{C} } (3-\frac{r^{2} }{R_{C}^{2} } ),{ \; \; \; \; \; \; \; }}
r\le R_{C}, \\ {\frac{Z_{1} Z_{2} e^{2} }{r}, { \; \; \; \; \; \;
\; \; \; \; \; \; \; \; \; \; \; \; \; \; }}
r>R_{C}, \end{array}\right.
\end{equation}
where $Z_1e$ and $Z_2e$ are the charges of the particles;
$R_C=r_C\,A^{1/3}$ ($r_C$ is the parameter of the Coulomb radius).

The radial wave function $u_l(r)$ for the partial wave with the
orbital momentum \textit{l} is the solution of the radial
Schr\"odinger equation ($\mu_{12}$ is the reduced mass, $E$ is the
energy in CM system)
\begin{equation} \label{GrindEQ__3_}
\left\{\frac{d^2}{dr^2} +2\mu_{12}
\left[E-V(r)\right]-\frac{l(l+1)}{r^2}\right\}u_{l} (r)=0.
\end{equation}
Here, $u_l(r)$ satisfies the standard boundary condition at the
origin:
\begin{equation} \label{GrindEQ__4_}
\left. u_{l} (r)\right|_{r=0} =0.
\end{equation}
To write the boundary condition for the derivative of $u_l(r)$, we
analyze the behavior of the wave function near the origin. The sum
of the potentials $V(r)=V_{N} (r)+V_{C} (r)$ multiplied by
\textit{r} is limited. Therefore we choose the point $r_{0} $ near
the origin, and denote $k_{0}^{2} =2\mu_{12} \left[E-V(r_{0}
)\right]$.

The solution of the Schr\"odinger equation
\begin{equation} \label{GrindEQ__5_} \left\{\frac{d^2}{dr^2} +k_{0}^{2} -
\frac{l(l+1)}{r^2} \right\}u_{l} (r)=0, \end{equation}
which satisfies the condition  (\ref{GrindEQ__4_}),  is proportional to
the function $g_{l} (k_{0} r)=k_{0} r\cdot j_{l} (k_{0} r)$, where
$j_{l} (k_{0} r)$ is the spherical Bessel function. Taking this
into account, one can use the initial condition for Eq.
(\ref{GrindEQ__3_}) as follows

\begin{equation} \label{GrindEQ__6_} \left. u_{l} (r)\right|_{r=r_{0} }
=g_{l} (k_{0} r_{0} ),{\rm \; \; \; \; \;  }\left. u'_{l}
(r)\right|_{r=r_{0} } =k_{0} g'_{l} (k_{0} r_{0} ).
\end{equation}
Note that the energy \textit{E} is negative for bound and virtual
states and complex for  resonance states. In the external region
$r>R_0$, where the nuclear potential can be omitted with
reasonable accuracy, the general solution of Eq.
(\ref{GrindEQ__3_}) is given by
\begin{equation}\label{GrindEQ__9_}
u_{l} (r)\cong u_{l}^{as} (r)=C_{l}^{(-)}(k)u_{l}^{(-)}
(kr)-C_{l}^{(+)} (k)u_{l}^{(+)} (kr),
\end{equation}
where $k=\sqrt{2\mu_{12} E} $, $C_{l}^{(\pm )} $ are the
coefficients that can be found by matching $u_{l}(r)$ to the
solution in the internal region at $r=R_{0}$  \footnote[1]{Note that in
contrast to the asymptotic wave function used in \cite{vertse82} our
asymptotic function (\ref{GrindEQ__9_}) contains both
outgoing wave and incoming wave.}. The functions $u_{l}^{(\pm )}
(\rho )$ can be written in terms of the regular $F_{l} (\eta ,\rho
)$ and the irregular $G_{l} (\eta ,\rho )$ Coulomb wave functions
\begin{equation} \label{GrindEQ__10_} u_{l}^{(\pm )} (\rho )=
e^{\mp \delta _{l}^{C} } \left[G_{l} (\eta ,\rho )\pm iF_{l} (\eta
,\rho )\right], \end{equation} where $\eta = Z_1Z_2e^2\mu_{12}/k$ is
the Sommerfeld parameter, $\delta _{l}^{C} $ is the Coulomb phase
shift given by $\delta _{l}^{C}=\arg\Gamma(1+l+i\eta)$ and $\rho
=kr$. The asymptotic forms of $u_{l}^{(\pm )} (\rho )$ are given
by
\begin{equation}\label{GrindEQ__11_}
u_{l}^{(+)} (\rho )\to \exp \left[i\left(\rho -\eta \ln 2\rho
-\frac{l\pi }{2} \right)\right],{\rm \; \; \; \; \; \;}\rho \to
\infty ,
\end{equation}
\begin{equation}\label{GrindEQ__12_}
u_{l}^{(-)} (\rho )\to \exp \left[-i\left(\rho -\eta \ln 2\rho
-\frac{l\pi }{2} \right)\right],{\rm \; \; \; \; \;  }\rho \to
\infty.
\end{equation}
 The coefficients
$C_{l}^{(+)} (k)$ and $C_{l}^{(-)} (k)$ are proportional to the
corresponding Jost functions \cite{mes66, new66}. The functions
(\ref{GrindEQ__11_}) and (\ref{GrindEQ__12_}) describe outgoing
and incoming waves, respectively. We can solve the Schr\"odinger
equation numerically and search for the energy at which the
coefficient $C_{l}^{(-)} (k)$ vanishes. This condition
($C_{l}^{(-)}(k)=0$) means that we are dealing with only the
outgoing wave in the asymptotic region ($r\to \infty $). Note that
for virtual and resonance states the first term in Eq.
(\ref{GrindEQ__9_}) is much smaller than the second one, which makes
it difficult to obtain a solution and an eigenvalue. To make sure
that $C_{l}^{(-)} (k)/C_{l}^{(+)} (k)$ goes to zero, we calculate
the ratio of the Schr\"odinger equation solution (for the sum of a
nuclear and the Coulomb potentials) and the outgoing wave in the
Coulomb potential. This ratio must approach a constant in the
asymptotic region. We also check the equality of the
logarithmic derivatives of $u_{l} (r)$ and $u_{l}^{(+)}(kr)$ at $r=R_0$.
$R_0$ should be chosen a little larger than the radius of the nuclear
potential. According to the scattering theory \cite{new66}, the vanishing
of $C_{l}^{(-)}$ at the positive imaginary semi-axis in the
complex momentum plane  corresponds to the bound state, while that
on the negative imaginary semi-axis corresponds to the virtual
(antibound) state. The resonant state is defined  by the zero of
$C_{l}^{(-)}$ in the fourth quadrant of the complex momentum
plane.

The $S$-matrix is the ratio $C_{l}^{(+)}(k)/C_{l}^{(-)}(k)$, which has a pole at
$k=k_{0}$ if $C_{l}^{(-)}(k_{0})=0$  \cite{mes66}.
For the poles of $S$-matrix of the first order the residue at the pole $k_0$ should be
\begin{equation}
{\rm{Res}}\,(S_l(k_0))=
A_l(k_0)=\frac{C_{l}^{(+)}(k_0)}{C_{l}^{(-)\prime}(k_0)},
\label{residue}
\end{equation}
where $C_{l}^{(-)\prime}(k_0)$ is the derivative at the pole
$k=k_0$. To find $A_l(k_0)$, we calculate $C_{l}^{(-)}(k)$  close
enough to the pole $k_0$. Then, we use the fit function
\begin{equation} C_{l}^{(-)}(k)=a_1(k-k_0)+a_2(k-k_0)^2,
\end{equation}
to obtain the coefficients of the expansion $a_1$ and $a_2$ for
which $C_{l}^{(-)\prime}(k_0)=a_1$. The described method we call the potential
$S$-matrix pole method.

\section{RESULTS}
\subsection{The bound states of ${}^{14}\rm{N}$}

To show how the method works, we start from its application to the
bound states of ${}^{14}\rm{N}$ considering it as a two-body bound
state ${}^{14} {\rm N}={}^{13}{\rm C}+p$. We assume that the
proton in the $1p_{1/2}$ orbital is coupled to the $1/2^{-}$
ground state of $^{13}\rm{C}$ to form the $1^{+}$ ground state and $0^{+}$
excited state of $^{14}\rm{N}$.

\begin{figure}[bp]
\resizebox*{0.48\textwidth}{!}{\includegraphics{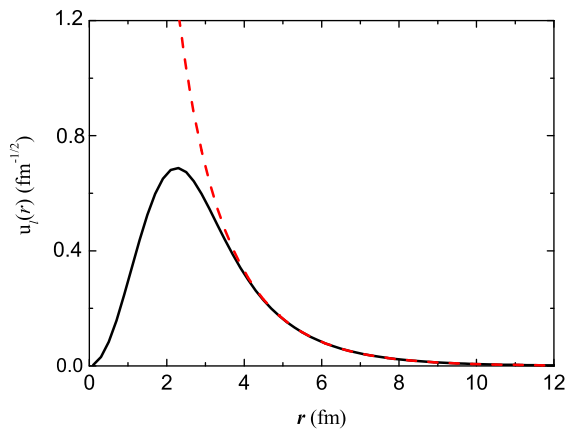}}
\caption{Comparison of the normalized radial bound state wave
function for the ${}^{14}{\rm{N}}\;\; (1^+)$ state (solid line)
with the corresponding asymptotic form ($W_{-\eta,l+1/2}(2\kappa
r)$, dashed line).} \label{fig_1}
\end{figure}

\begin{figure}[bp]
\resizebox*{0.48\textwidth}{!}{\includegraphics{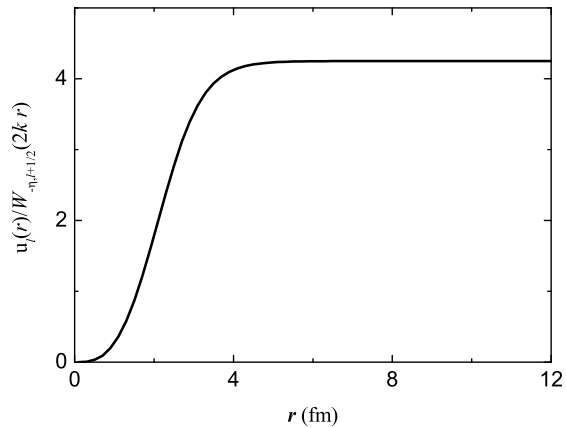}}
\caption{Ratio of the calculated radial bound state wave function
to the Whittaker function ($W_{-\eta,l+1/2}(2\kappa r)$) for the
${}^{14}\rm{N}\;\; (1^+)$ state.} \label{fig_2}
\end{figure}

To describe these states in the two-body (core + nucleon) approach,
we choose the geometrical parameters of the bound state
Woods-Saxon potential to be $r_0=r_C=1.2$ fm and $a=0.5$ fm. The
well-depth procedure providing the experimental binding energy
leads to $V_0=51.65$ MeV and $V_{LS}=1.5$ MeV  for the $1^{+}$
state and $V_0=47.71$ MeV and $\,V_{LS}=1.5$ MeV for the $0^{+}$
state. The coefficients $C_{l}^{(+)}(k)$ and $C_{l}^{(-)}(k)$
are found from the set of equations $u_{l} (r_{1} ){\rm \;
}=u_{l}^{as}(r_{1})$, $u_{l}(r_{2}){\rm \;}=u_{l}^{as}(r_{2})$
($u_{l}^{as}$ is the known asymptotic solution), where both the
neighboring points $r_1$ and $r_2$ should be chosen in the
asymptotic region. In this work we choose as an example $r_1  =
0.5 R_{max}$ and $r_2 = 0.51 R_{max}$, where $R_{max }=N\,R_N$.
The parameter \textit{N} should be big enough to fulfill the
condition $u_{l}(r_{1})/u_{l}^{as}(r_{1})$= \textit{const}. In
Figs. \ref{fig_1} and \ref{fig_2}, the wave function for the $1^+$ state of
${}^{14}\rm{N}$ and the ratio of the wave function to the
Whittaker function describing its asymptotic behavior are shown.
For the $0^+$ state, the wave function and its ratio to the
Whittaker function are very similar to those of the $1^+$ state.
From these figures, one can conclude that the coefficient
$C_{l}^{(-)}(k)$ is equal to zero and the wave function coincides
with its asymptotic form when $r > R_{0}$.

From Eq. (\ref{GrindEQ__9_}) we get that in the external region
the radial bound state wave function is given by
\begin{equation}
 u_{l}^{(bs)as} (r)=C_{l}^{(+)}(k)u_{l}^{(+)}(kr),
\label{bswf1}
\end{equation}
where $k=i\kappa_{bs}\,\,(\kappa_{bs}>0)$. Normalizing the bound
state wave function to unity we can rewrite its asymptotic term as
\begin{equation}
 u_{l}^{(bs)as} (r)= b_{l}\,W_{-\eta_{bs}, l+1/2}(2\,\kappa_{bs}\,r),
\label{bswf2}
\end{equation}
where $b_{l}$ is the single-particle ANC, $W_{-\eta_{bs},
l+1/2}(2\,\kappa_{bs}\,r)$ is the Whittaker function determining
the radial shape of the bound state wave function, $\eta_{bs}$ is
the Coulomb parameter for the bound state,
$\kappa_{bs}=\sqrt{2\,\mu_{12}\varepsilon_{bs}}$ is the bound state
wave number and $\varepsilon_{bs}$ is the binding energy of the
bound state. For the adopted geometrical parameters, we get
$b_{1(gr)}= 4.250\,\, \rm{fm}^{-1/2}$ for the ground state and
$b_{1(exc)} = 2.457\,\,\rm{fm}^{-1/2}$ for the excited state.
Note that the single-particle ANC is sensitive to these parameters
\cite{akram03}.

The residue at the bound state pole of the $S$-matrix is given by \cite{baz}
\begin{equation}\label{W-ANC}
A_l(k_0)=(-1)^{l+1}i\,b_l^2.
\end{equation}
Our calculated residues of the $S$-matrix at poles related to the
ground and excited states give the values
$A_{gr}=18.061\,i\,\,\rm{fm}^{-1}$ and
$A_{exc}=6.039\,i\,\,\rm{fm}^{-1}$, respectively.  Found from these
residues, the single-particle ANCs coincide with $b_{1(gr)}$ and
$b_{1(exc)}$ given above and found from the bound state wave
functions. This validates method of calculation of
the residue of the $S$-matrix at the bound state pole presented here.

\subsection{Virtual (antibound) state}

   Here we apply our method to obtain the energy of the virtual
(antibound) state in the $n\,p$ system at $l=0$, taking into account only
the short-range Yukawa nuclear potential  $V_N(r) = V_{0}\,r^{-1}\,\exp(-r/R)$.
The virtual state corresponds to $k=-i\,\kappa$ $(\kappa>0)$,
i.e. the pole of the $S$-matrix is located on the negative imaginary
semi-axis in the complex momentum plane. It generates the exponentially
increasing term $u_{0}^{(+)}(kr)$ when $r \to \infty$ while the second term
$u_{0}^{(-)}(kr)$ becomes exponentially small making it very
difficult to determine the energy (momentum) when $C_{0}^{(-)}=0$,
which is the condition for the virtual pole. For this reason, to
calculate $C_{0}^{(-)}(-i\kappa )$, one should obtain a solution
with very high precision. In our calculations, the energy $\mid
\epsilon_{v}(N)\mid$ of the virtual state calculated as function
of $R_{max }=N\,R_N$ decreases smoothly as
\textit{N} increases. However, when $N>N_{max}$ the
energy exhibits a sudden change to the larger value. It means that
for $r \geq R_{max }$ the solution is not precise enough to
calculate $C_{0}^{(-)}(-i\kappa )$ accurately. That is why we
adopt $\epsilon_{np}=\epsilon_{v}(N_{max})$ as the virtual pole
energy. Our result $\epsilon_{np}=-0.067$ MeV agrees very well
with the one obtained using the integral equation method \cite{irg06, orl200}.
The calculated residue of $S$-matrix in pole is $A_{np}=-0.072i\,\,\rm{fm}^{-1}$
leading to the single-particle ANC for the virtual $n\,p$ state
$b_{0}=0.268$ fm${}^{-1/2}$.

\subsection{The resonance states of $^{15}\rm{F}(1/2^+,\, 5/2^+)$}

Several articles were published recently \cite{baye2005, qi2008,
fortune2006, canton2006, canton2007} testing the predictive power
of the current theoretical approaches to describe the lowest broad
levels in ${}^{15}{\rm F}$. The final goal of these analyses was a
comparison of the predictions with the available experimental data
on the ${}^{15}{\rm F}$ levels. Determination of  a broad  resonance
parameters is a well known unsolved problem in physics.
The resonance energy and width for a broad
resonance are not defined uniquely and there are many prescriptions, which have been used in literature \cite{barker96}. The definitions depend not only on the model used, say potential, $R$-matrix, microscopic, but even within a given model the prescriptions for the resonance parameters can be different \cite{barker87,barker96}. For example, in \cite{sherrbertsch} four different definitions were used. In \cite{barker96} two more definitions were added in the $R$-matrix approach. That is why we believe that,  when any compilation includes the broad resonance parameters, the reference should be done to the prescriptions used to determine these parameters. The reason for this ambiguity is that for broad resonances in the physical region  the nonresonant contribution becomes comparable with the resonant one.
In this case the determined resonance energy and width depend  on how much of  the  background  is included into the resonant part. The only way to determine correctly
the resonance energy and width is to single out the resonance pole explicitly in the function fitting the experimental data. It is realized in the $S$-matrix pole method.

 Here we address two approaches based on the definition of the resonance energy $E_{R}= E_{0} - i\,\Gamma/2$ as the energy at which the $S$-matrix has a pole on the second energy sheet (low half of the momentum plane)
: the potential approach based on the solution of the radial Schr\"odinger equation and the analytical expression for the $S$-matrix. The first one gives the most accurate definition of the resonance energy and width within the potential model, while the second one even more general because it based only on the analyticity and the symmetry of the $S$-matrix \cite{baz}.

We remind that a resonance
corresponds to the pole of the $S$-matrix at $k_{R} = k_{0} -
i\,k_{I}$ and is located in the fourth quadrant of the momentum
complex plane. Correspondingly the resonance energy is
\begin{equation}
E_{R}= \frac{k_{R}^{2}}{2\,\mu}= E_{0} - i\,\frac{\Gamma}{2},
\label{resenergy1}
\end{equation}
where
\begin{equation}
E_{0}= \frac{k_{0}^{2} - k_{I}^{2}}{2\,\mu}, \label{realresenergy1}
\end{equation}
and
\begin{equation}
\Gamma= \frac{ 2\,k_{0}\,k_{I} }{\mu}. \label{reswidth1}
\end{equation}
For broad resonances $k_{I}$ becomes comparable with $k_{0}$ or even larger 
($k_{0} \lesssim k_{I} $). If $k_{I} > k_{0}$, i.e. the resonant pole
in the complex momentum plane, due to large $k_{I}$, is far from
the real energy axis and the energy of the broad 
resonance, $\,E_{0} <0$, is located in the
third quadrant on the second energy sheet and we call it the 
subthreshold broad resonance \footnote[2]{In literature another definition of the subthreshold resonance is also being used: the resonance is subthreshold if in the resonance
reaction $\alpha \to \beta$ the resonance energy in the initial channel is negative.}. 
Due to large  $k_{I}$ (or resonance width $\Gamma$), the impact of the resonant pole on
the cross section or scattering phase shift is weakened and the
non-resonant amplitude or phase shift (non-resonant background)
becomes important. The general expression for the elastic
scattering $S$-matrix based on its analyticity and symmetry in a
vicinity  of a single resonance can be written as  \cite{baz}
\begin{align}
S(k) = e^{2\,i\,\delta(k)}= e^{2i\,\delta_{p}(k)}\,\frac{(k -
k_{R}^{*}) (k + k_{R})}{(k - k_{R})(k + k_{R}^*)}
\nonumber\\
=e^{2\,i\,(\delta_{p}(k) + \delta_{R}(k) + \delta_{a}(k))}.
\label{elsmatr1}
\end{align}
where  $\delta_{p}(k)$ is the non-resonant scattering
phase shift,
\begin{align}
\delta_{R}(k) = -\arctan{\frac{k_{I}}{k - k_{0}}} \label{deltaR1} \\
=-[\frac{\pi}{2} - \arctan{\frac{k - k_{0}}{k_{I}}}] , \label{deltaR2}
\end{align}
is the resonant scattering phase shift
\footnote[3]{Note that Eq. (\ref{deltaR1}) is valid for any $k_{I}$
for $ k - k_{0} \geq 0 $ while Eq. (\ref{deltaR2}) is valid
for $k_{I} >0$  and any $k - k_{0}$.}, and
\begin{equation}
\delta_{a}(k) =-\arctan{\frac{k_{I}}{k + k_{0}}}. \label{deltaa1}
\end{equation}
For narrow resonances, $k_{I} << k_{0}$, the phase shift
$|\delta_{a}(k)| << 1$ can be neglected. In this case, the
standard method, which we call the phase shift method (or
"$\delta=\pi/2$" rule), entails the resonance energy $E_0$ the
value at which the scattering phase $\delta(k)$ passes through
$\pi/2$. The resonant width is evaluated from the formula
$\Gamma=2/(d\delta/dE)$ at $E=E_0$ or as the energy interval
corresponding to change of $\delta$ from $\pi/4$ to $3\pi/4$.
However, for broad resonances $\delta_{a}(k)$ cannot be neglected
and the total non-resonant scattering phase shift $\delta_{p}(k) +
\delta_{a}(k)$ becomes dependent on the resonant parameters. This
non-resonant scattering phase shift may be a large negative so that
the total phase shift $\delta(k)$ cannot reach $\pi/2$ at $k =
k_{0}$ making the $\pi/2$ method non-applicable. When calculating
the elastic cross section or scattering phase shift in the
presence of the broad resonance, due to the importance of the
non-resonant phase shift, the cross section depends not only on
the resonance parameters $E_{0}$ and $\Gamma$ but also on the
potential adopted.

Here as a test case we select resonances representing the ground
state $1/2^{+}$ and the first exited state $5/2^{+}$ in
${}^{15}{\rm F}$. The latest very detailed analysis of  the
angular distributions for the ${}^{14}{\rm C}(d,p){}^{15}{\rm C}$
reaction \cite{pang2007, murillo1994} showes that the
spectroscopic factors for the ground $1/2^{+}$ and the first
excited state $5/2^{+}$ are close to the single particle ones
(0.99 and 0.90 correspondingly \cite{murillo1994}. One expects
the similar numbers for the mirror states in ${}^{15}{\rm
F}$. Therefore, the potential approach is appropriate to describe these
states. In \cite{gol04} the Woods-Saxon potential parameters to describe the
excitation energies of the mirror levels in ${}^{15}{\rm C}$ and
${}^{15}{\rm F}$ as well as the experimental data on resonance
${}^{14}{\rm O}+p$ were found. The authors
\cite{gol04} presented the final data on the resonance parameters
for the first two levels in ${}^{15}{\rm F}$ using the
calculations of the wave function inside the nucleus, at the
radius of 1 fm. The energy at which the absolute value of the
wave function reaches its maximum was identified as the resonance
energy. We call this the $|\Psi_{max}|$ method. In \cite{gol04} the width of the
resonance was defined by the energy interval over which the amplitude falls
by $\sqrt{2}$ relative to the maximum of the $|\Psi_{max}|$. For comparison, in
\cite{gol04} some results were presented using also the $\pi/2$ method.

We apply the potential $S$-matrix pole method by solving the Schr\"odinger
equation with the Woods-Saxon potential given in \cite{gol04} for both ${}^{15}\rm{F}$ resonance states with the $J^{\pi} = 1/2^+$ and $5/2^+$. We search for the complex
energy at which the coefficient $C_l^{(-)}=0$ (see Eq. (\ref{GrindEQ__9_}))
similar to the search for the bound or the virtual state.
We note that in the standard approach the scattering wave function
is calculated at real energies, where the non-resonant
contribution is significant for  broad resonances, while the Gamow
wave function is calculated at the complex energy corresponding to
the resonant pole of the $S$-matrix located on the second Riemann
energy sheet. As a first approximation, to determine the
complex resonance energy $E_R^{1}=E_{0}^{1} -i\Gamma^{1}/2$ we use the phase shift
method (or the $|\Psi_{max}|$ method when the $\delta=\pi/2$ method
is non-applicable). After that, we solve the Schr\"odinger equation near the complex
energy $E_R^{1}=E_0^{1}-i\Gamma^{1}/2$. The final result of this search is the
complex energy $E_R$, at which the coefficient of the incoming
wave vanishes. We also applied the $S$-matrix pole search using the analytical
representation (\ref{elsmatr1}) for the $S$-matrix (see explanation below).

Our results for the energies and widths of the resonance states
are given in Table \ref{table1} compared with the previous
results obtained using the $\delta=\pi/2$ and $|\Psi_{max}|$ methods \cite{gol04}.
The position \textit{E} and the width $\Gamma$ of
the broad resonance depend on the calculation method: the $S$-matrix
pole method gives the values of the resonance energy and width
smaller and more accurate than the $\delta=\pi/2$ and
$|\Psi_{max}|$ methods.
\begin{table}
\caption{ Energy and width of the resonances for the
${}^{15}\rm{F}$ states with $J^p = 1/2^+$ (the ground state) and
$5/2^+$ (the first exited state) calculated by the use of three
different methods (see the text) \label{table1}}.
\begin{ruledtabular}
\begin{tabular}{clll}
$J^p$ & $E_0$ (MeV) & $\Gamma$ (MeV) & Method \\ \cline{1-4}
& 1.450 &  1.091 & $\delta=\pi/2$ \\
         & $1.290_{-0.06}^{+0.08}$ &  0.7 & $|\Psi_{max}|$ \newline  \\
         & 1.198 &  0.530 & Pole of $S$-matrix (potential)   \\
$1/2^+$  & 1.194 &  0.531 & Pole of $S$-matrix using Eq. (\ref{elsmatr1})   \\
         & 1.400 &  0.700 & $R$-matrix (from\\
         &       &        &  the scattering phase shift)   \\
         & 1.315 &  0.679 & $R$-matrix (from the excitation \\
         &       &        &  function, $r_{0}=4.5$ fm)   \\
         & 1.274 &  0.510 & $R$-matrix (from the excitation \\
         &       &        &  function, $r_{0}=6.0$ fm)   \\\hline
         & 2.805 & 0.304 & $\delta=\pi/2$ \\
         & $ 2.795\pm  0.045$ & $ 0.298 \pm  0.06 $\footnotemark[1] & $|\Psi_{max}|$ \\
$5/2^+$  & 2.780 &  0.293 & Pole of $S$-matrix \\
         & 2.777 &  0.286 & $R$-matrix (from the excitation \\
         &       &        & function, $r_{0}=4.5$ fm) \\
         & 2.762 &  0.297 & $R$-matrix (from the excitation \\
         &       &        & function, $r_{0}=6.0$ fm) \\
\end{tabular}
\footnotetext[1]{\, It was misprint $\Gamma=0.325$ MeV for the
state $\frac{5}{2}^+$ in Ref. \cite{gol04}.}
\end{ruledtabular}
\end{table}
It is worth noting that the corrected value of $1.227$ MeV for the resonance energy  of the ground state of ${}^{15}{\rm F}$ is very close to the lower limit given by Fortune \cite{fortune2006} obtained using the isobaric multiplet mass equation. Besides, in the most recent experimental work on ${}^{15}{\rm F}$ \cite{guo2005} it was indicated that the ground state energy of ${}^{15}{\rm F}$ can be even lower.

\begin{figure}[bp]
\resizebox*{0.48\textwidth}{!}{\includegraphics{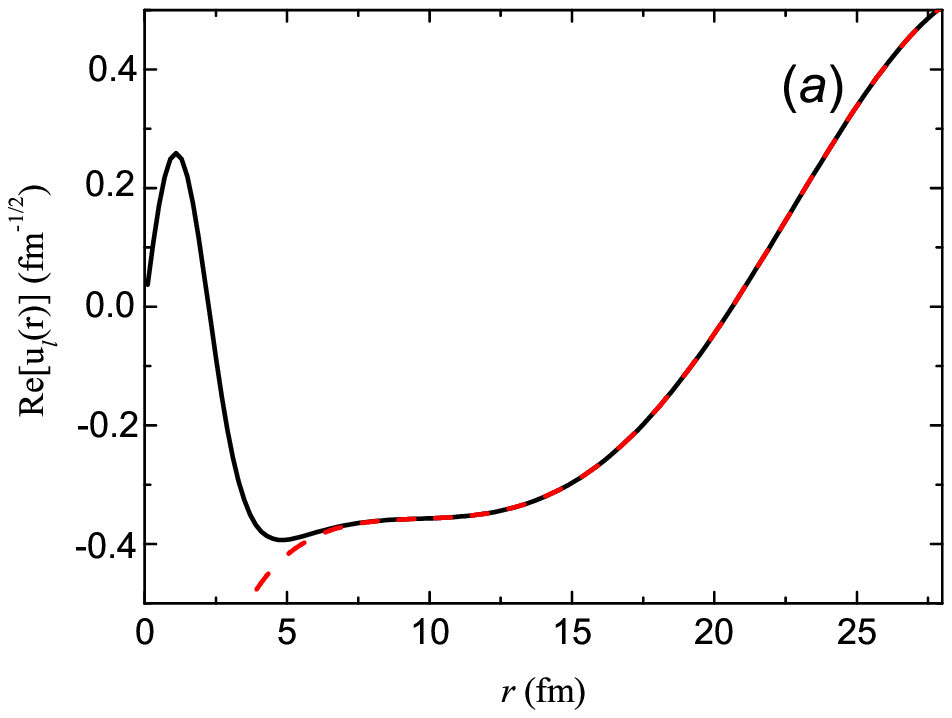}}
\resizebox*{0.48\textwidth}{!}{\includegraphics{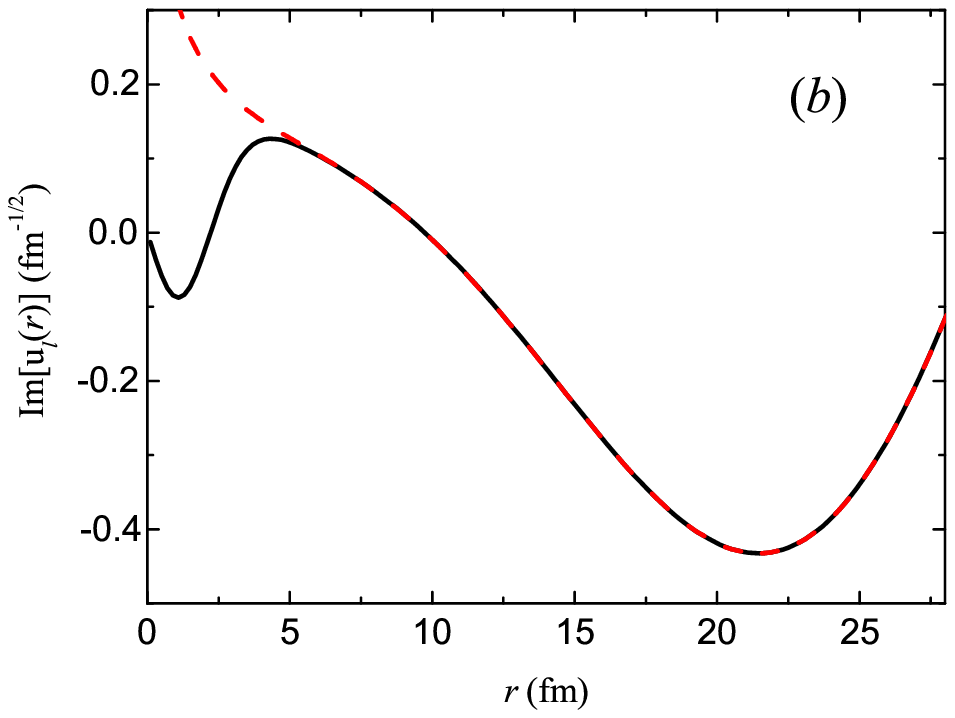}}
\caption{Real ({\it a}) and imaginary ({\it b}) parts of the wave
function of the $1/2^+$ resonance state in ${}^{15}\rm{F}$. The
solid line is the solution of the Schrödinger equation, the dashed
line is the outgoing Coulomb function (the Whittaker function).}
\label{fig_3}
\end{figure}

\begin{figure}[bp]
\resizebox*{0.48\textwidth}{!}{\includegraphics{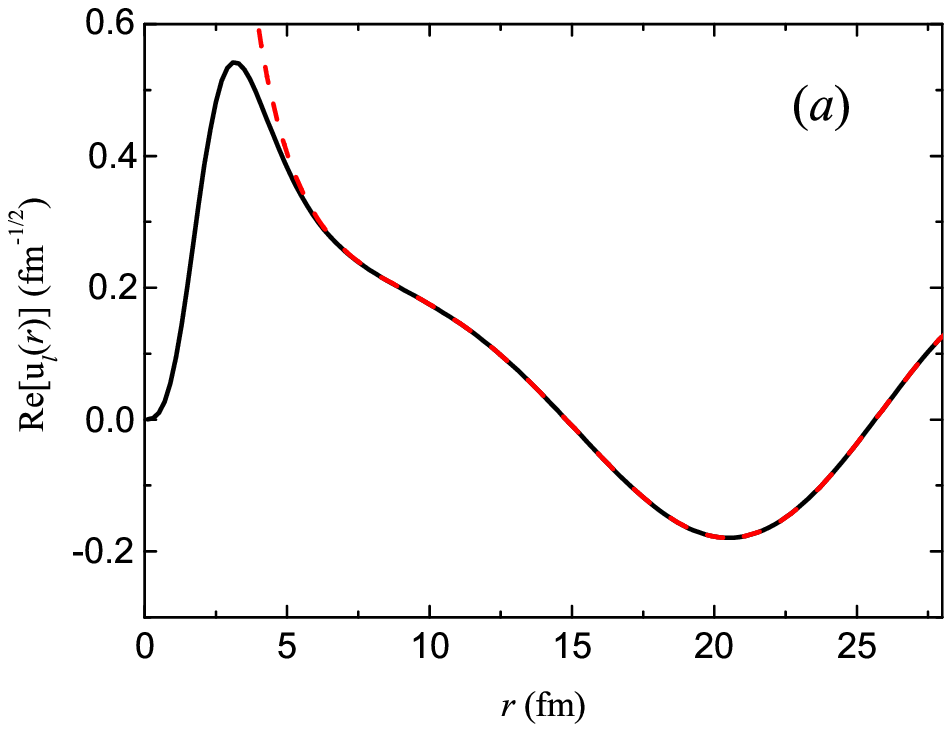}}
\resizebox*{0.48\textwidth}{!}{\includegraphics{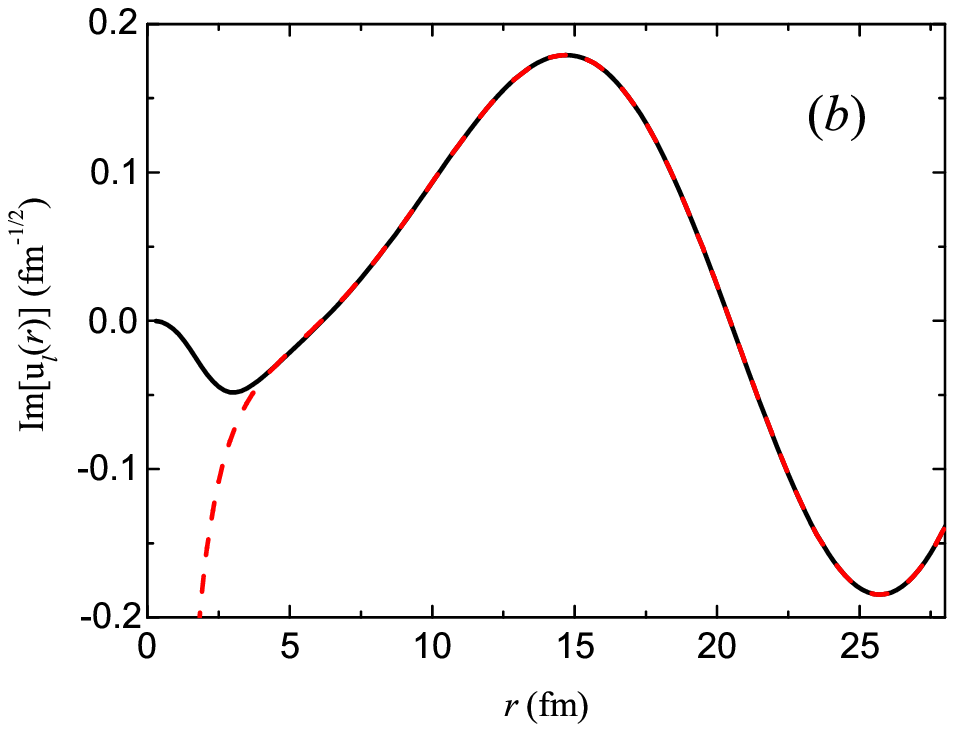}}
\caption{The same as in Fig.~\ref{fig_3} but for the $5/2^+$
resonance state in ${}^{15}\rm{F}$.} \label{fig_4}
\end{figure}

Figs. \ref{fig_3} and \ref{fig_4} show the real and imaginary parts of
the normalized Gamow wave function for the $1/2^+$ and
$5/2^+$ resonance states in ${}^{15}\rm{F}$. The solution of the Schr\"odinger
equation coincides with the outgoing wave outside the potential area.
We conclude that the probability of finding the proton inside the
potential area is relatively high. The advantage of our
method is that we directly find the complex energy of the resonant
state (energy and width) by the same procedure as for the bound
state.

An important test of our method is comparison of the single-particle
ANC determined as an amplitude of the tail of the normalized Gamow
function with the ANC  determined from the residue of the scattering
amplitude at the pole corresponding to the resonance.
For the normalization of the Gamow wave function we use here the method
suggested by Zeldovich \cite{zel60}, the numerical application of which is
difficult for a broad resonance. However, the same relationship between the
squared single-particle ANC and the residue can be used for both
the bound and resonance states. One can use Eq. (\ref{W-ANC})
to find the single-particle ANC of the resonance wave function.
The results of the calculated residues are $(-0.038 + i\,0.008)
\,\,\rm{fm}^{-1}$ and $(0.015 - i\,0.009) \,\,\rm{fm}^{-1}$ for
the $1/2^+$ and $5/2^+$ states, respectively. From Eq.
(\ref{W-ANC}) we get the single-particle ANCs $(-0.123 + i\,0.153)
\,\,\rm{fm}^{-1/2}$ and $(0.115 + i\,0.067) \,\,\rm{fm}^{-1/2}$
for the same states, correspondingly. We obtained the same
single-particle ANCs directly from the tail of the normalized Gamow
wave functions validating Eq. (\ref{W-ANC}).

\subsubsection{Model-independent determination of the energy and width of the broad resonance $\frac{1}{2}^{+}$ in $^{15}\rm{F}$}

\begin{figure}[bp]
\resizebox*{0.48\textwidth}{!}{\includegraphics{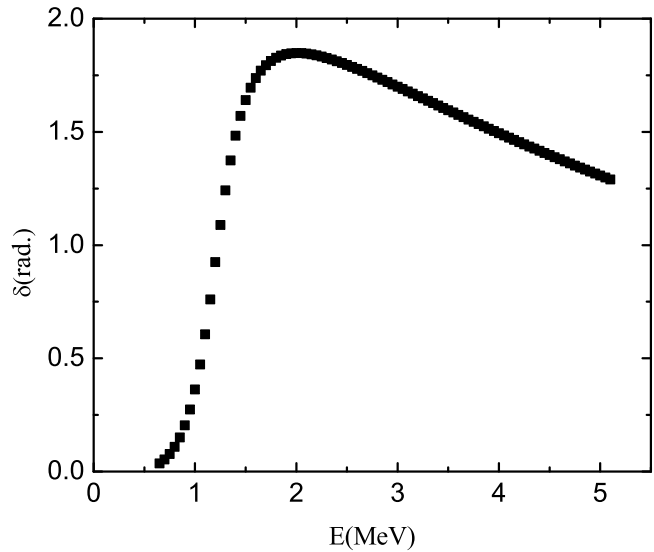}}
\caption{The ${}^{14}{\rm O}+ p$ $\,2\,s_{1/2}$ scattering phase shift generated
by the Woods-Saxon potential from \cite{gol04} and used as the "quasi-experimental phase
shift".}
\label{fig_phashift15F1}
\end{figure}

The limitations of the potential model and the existence of the phase-equivalent potentials calls for a cross check of the energy and width for the broad resonance determined from the potential approach. We demonstrate how to determine these resonance parameters using the model-independent representation of the elastic scattering $S$-matrix given by Eq. (\ref{elsmatr1}).  Since the experimental $2s_{1/2}$ phase shift for ${}^{10}{\rm C} +p$ scattering in the resonance energy region is not available, we generate the "quasi-experimental"
$2s_{1/2}$ phase shift using the Woods-Saxon potential from \cite{gol04}, which reproduces
the ${}^{14}{\rm O} + p$ resonance scattering. Its geometry is $r_{0}=1.17$ fm, $a=0.735$ fm,
$r_{C}=1.21$ fm and the depth $V_{0} = 53.52$ MeV. 
The phase shift is shown in Fig \ref{fig_phashift15F1}.
Using the $S$-matrix pole method from the solution of the Schr\"odinger equation we find the resonance energy for this potential $E_{0}= 1.198$ MeV and the resonance width $\Gamma= 0.530$ MeV. Now we demonstrate that using Eq. (\ref{elsmatr1}) we can fit the
"quasi-experimental" phase shift and determine the resonance energy and width.
The potential phase shift in Eq. (\ref{elsmatr1}) is approximated by the polynomial $\delta_{p}(k)= \sum\limits_{n = 0}^3 {{b_n}{{(k - {k_s})}^n}} $. So, we have 6 fitting parameters including 4 coefficients $b_{n}$, $\,E_{0}$ and $\Gamma$. The final result does not depend on the choice of the center of the Taylor expansion $k_{s}$ and practically not sensitive to the starting values of $E_{0}$ and $\Gamma$.
We take here the starting values $k_{s}= 0.25$ fm$^{-1},$ $\,E_{0}=1.45$ MeV and $\Gamma= 1.276$ MeV obtained from the $\delta= \pi/2$ method, Table \ref{table1}. The fit to the "qausiexperimental" phase shift gives the final resonance energy
$E_{0}= 1.194$ MeV and $\Gamma= 0.531$ MeV what is in a perfect agreement with the results
obtained using the potential $S$-matrix pole method.  For the starting search values $E_{0}=1.6$ Mev and $\Gamma=1.276$ MeV we get the fitted energy $E_{0}= 1.198$ MeV and
$\Gamma= 0.532$ MeV. Thus Eq. (\ref{elsmatr1}) allows one to obtain the energy and width of the broad resonance using, for example, as input parameters the resonance and width obtained by the $\delta=\pi/2$, $\,|\Psi_{max}|$. The model-independent result obtained from Eq. (\ref{elsmatr1}) gives very close values to the potential $S$-matrix pole. Assigning  a $10\%$ uncertainty to the "quasi-experimental" phase shift results in a similar uncertainty in the determined "quasi-experimental" phase shift  resonance energy and width.

\subsubsection{Comparison with $R$-matrix approach}

The resonant  $S$-matrix obtain from the $R$-matrix contains the nonresonant contribution through the energy dependence of the level shift and resonance width. The extrapolation of this functions to the complex energy plane make them complex, i.e. they lose it physical meaning. Thus the $R$-matrix approach is not designed for extrapolation to the resonant pole.

Here we apply the $R$-matrix approach to determine the energy and the width of the resonance with the $S$-matrix pole method. For an isolated resonance in the single-level,
single-channel $R$-matrix approach with the zero boundary condition the Coulomb-modified nuclear scattering $S$-matrix is
\begin{equation}
S=e^{2\,i\,\delta_{hs}}\,\frac{E_{0} - E + i\,\frac{\Gamma(E)}{2}}{ E_{0} - E - i\,\frac{\Gamma(E)}{2}},
\label{rmatrampl1}
\end{equation}
where $\delta_{hs}$ is the hard-sphere scattering phase shift. To obtain this equation the linear energy dependence of the level shift function $\Delta(E)$ is taken into account \cite{lanethomas}. Here, $E_{0}$ is the real part of the resonance energy. In the $R$-matrix $E_{0}$ is determined as $E_{\lambda} +  \Delta(E_{0})=E_{0}$,  $\,E_{\lambda}$ 
is the $R$-matrix level energy,  $\,\Gamma(E)= 2\,\gamma_{l}^{2}\,P_{l}^{2}(E,r_{0})$ is the observable resonance width in the $R$-matrix approach depending on the energy and the channel radius $r_{0}$, $\,\gamma_{l}$ is the observable reduced width amplitude, $P_{l}(E,r_{0})$ is the penetrability factor in the $l$-th partial wave. The resonance width in the $R$-matrix approach, in contrast to the Breit-Wigner equation, depends on the energy. This dependence reflects the fact that the $S$-matrix in the $R$-matrix is richer than the Breit-Wigner equation: it includes also the non-resonant background,
which is contributed by the hard-sphere phase shift and the energy dependence of the level  shift function and the resonance width. For narrow resonance ($\Gamma(E_{0}) << E_{0}$) the pole in Eq. (\ref{rmatrampl1}) $E_{R} \approx E_{0} - i\,\Gamma(E_{0})/2$.
For a broad resonance this resonance energy is not a pole of the $S$-matrix. The equation for the resonant pole in this case is given by $E_{R} = E_{0} - i\,\Gamma(E_{R})/2$. At complex $E_{R}$ $\,\,\Gamma(E_{R})$ becomes complex and loses its meaning of the width.
For a broad resonance in the $R$-matrix method the resonance energy is defined as $E_{R}= E_{0} - i\,\Gamma(E_{0})$, which is not a pole of Eq. (\ref{rmatrampl1}). Hence, for broad resonances the difference between the resonance energy from the $S$-matrix pole method and the $R$-matrix method is expected.

To compare the results for the $R$-matrix and $S$-matrix pole
methods for the ${s_{1/2}}^{+}$ resonance we use the phase shift
generated by the Woods-Saxon potential from \cite{gol04} as the
"quasi-experimental" one and determine the resonance energy and
width by fitting the $R$-matrix phase shift to the
"quasi-experimental". The results are shown in Table \ref{table1}.
The $R$-matrix resonance energy and width found at $r_{0}=5.0$ fm
are higher than the $S$-matrix pole ones and close to the
$|\Psi_{max}|$ result. Both $R$-matrix and $|\Psi_{max}|$ methods
determine the resonance energy from the data at real energies
where for broad resonances the contribution of the background
becomes important. The $S$-matrix pole method determines the
resonance energy and width by extrapolating the data to the pole
in the complex energy (momentum) plane. In the vicinity of the
pole the resonant contribution becomes dominant compared to the
background and determination of the resonance parameters is more
accurate than in the physical region.

We made additional comparison of the $R$-matrix approach by fitting the measured in \cite{gol04}
the excitation function of the ${}^{14}{\rm O} + p$ scattering at $180^{\circ}$. Both resonances ${s_{1/2}}^{+}$ and ${d_{5/2}}^{+}$ coherently contribute to the excitation function. The resonances can be separated only after integration over the scattering angle. The selection of $180^{\circ}$ scattering angles minimizes the Coulomb scattering effects and enhances the $d_{5/2}^{+}$ resonance contribution. The two-level $R$-matrix fitting to the excitation function
gives the observable resonance energy and width presented in Table \ref{table1} for two channel radii $r_{0}=4.5$ and $6$ fm. The resonance energy is determined as the peak of the $|S(k)-1|^{2}$, and the width as the FWHM of this function. We note that this prescription differs from two prescriptions used in \cite{barker87}. For narrow ${d_{5/2}}^{+}$ all methods gives very close results, but it is not the case for the broad resonance ${s_{1/2}}^{+}$. The $R$-matrix results are between the $|\Psi_{max}|$ and the $S$-matrix pole. Since the $S$-matrix pole method
based on Eq. (\ref{elsmatr1}) correctly takes into account the resonance contribution as a pole in the complex energy (momentum) plane and analytically continue it to the physical region, it allows one to separate correctly the non-resonant (background) contribution from the resonance one and, hence, provides the most accurate determination of the resonance energy and width.

\subsection{The lowest levels in the mirror nuclei $^{11}\rm{Be}$ and $^{11}\rm{N}$}

The light neutron rich nucleus $^{11}\rm{Be}$ is probably the most
discussed nucleus. The interest to $^{11}\rm{Be}$ is related to the
well known inversion of the shell model levels in this nucleus. It has
the following low-lying states: $\frac{1}{2}^{+}$ (ground state),
and the excited states $\frac{1}{2}^{-}$  at $E_x=0.320$ MeV and
$\frac{5}{2}^{+}$ at $E_x=1.778$ MeV \cite{tunl}. The first two
are the bound states while the third one is a resonance.
As it was mentioned in \cite{mil2001} "the lowering of the $s_{1/2}$
orbital with respect to the $0d_{5/2}$ orbital is expected for a simple
potential well". The $p_{1/2}$ state belonging to the $K=1/2$ band
has a pretty stable dominantly $[421]$ spatial symmetry configuration
since the next $1/2^{-}$ state is $10$ MeV away \cite{millenerprivate}.
In \cite{kurathpicman} was shown that that the lowest $p_{1/2}$ state
obtained in a central potential with the spin-orbital interaction strongly
overlaps with the state projected from a Slater determinant of the lowest
orbits in the Nilsson's model with the same spin-orbital interaction as the shell
model and deformation.

In this work to test our method we apply it for calculation of the
three lowest states $s_{1/2},\,p_{1/2}$ and $d_{5/2}$ in
${}^{11}{\rm Be}$ and ${}^{11}{\rm N}$ nuclei belonging to the multiplet
$T=3/2$. We also estimate the spectroscopic factors for $s_{1/2}$ and
$d_{5/2}$ states using the potential approach leaving aside $p_{1/2}$
state, which is not a single-particle \cite{mil2001}.

Different reactions, including the $^{10}\rm{Be}(d,p){}^{11}{\rm Be}$
reaction with the radioactive $^{10}\rm{Be}$ target, were
used to obtain the spectroscopic factors for the lowest
states in $^{11}\rm{Be}$. As a standard  procedure, the single-particle
neutron wave functions in ${}^{11}{\rm Be}$ are used as the input in the  DWBA
code to get the neutron spectroscopic factors. The obtained spectroscopic factors are
in the intervals (0.5-0.96) \cite{sherr01} and (0.7-0.8) \cite{lima07}. But they are model-dependent because they depend on the Woods-Saxon potential adopted for the neutron bound state in ${}^{11}{\rm Be}$, optical potential in the initial and final channels of the $(d,p)$ reaction  and accuracy of the DWBA to analyze for the deformed
${}^{11}{\rm Be}$ \cite{mukh1997}. A priori the transfer reactions involving
deformed nuclei require the codes, which take into account the multi-step transfer
mechanisms, for example, the coupled channels Born approximation available in FRESCO.
That is why it is difficult to say from the DWBA analysis to what
extent the three lowest neutron states are single-particle.

The nucleus $^{11}\rm{N}$ is the mirror of $^{11}\rm{Be}$, and it
should have a similar level scheme. All states $^{11}\rm{N}$ are
unstable to proton decay. Therefore, their decay widths directly
related to their single particle nature. Since the discovery of
the ground state in $^{11}\rm{N}$ in 1996 \cite{axel96}, the
lowest levels in $^{11}\rm{N}$ were studied in many works (see
\cite{cas06} and references therein). In this section, we
apply the $S$-matrix pole method to study the broad
levels in $^{11}\rm{N}$. Simultaneously we attempt to find
restrictions on the single particle potentials related
to the widths and excitation energies of the mirror
states in $^{11}\rm{Be}$ and $^{11}\rm{N}$.

To determine the single-particles levels in $^{11}\rm{Be}$ and
$^{11}\rm{N}$, we use the Woods-Saxon plus Coulomb potential
similar to the ones used in \cite{fortune95,
barker96,sherr01,barker04}. The parameters of the potential are
fitted to reproduce the energies of the low-lying levels in
$^{11}\rm{Be}$. Then we use this nuclear potential plus the
Coulomb potential to find the mirror levels in $^{11}\rm{N}$.
We apply the pure single-particle approach as in
\cite{fortune95,barker96}.

The different sets of the potential parameters, which were used to
fit the lowest levels in $^{11}\rm{Be}$, are presented in Table
\ref{table2}. As a starting point, the standard geometrical
parameters $r_0=1.25$ fm, $a=0.65$ fm of the Woods-Saxon potential
are used. Then, we vary the depth of the central potential $V_{0}$
to fit the binding energy of the ground state $s_{1/2}$ of
${}^{11}{\rm Be}$ (well-depth procedure). After that, we vary the
radius $r_0$ and the diffuseness parameter $a$ to fit the binding
energy in $^{11}\rm{Be}$ at the fixed depth $V_0 = 57.057$ MeV
found from the fitting at standard geometrical parameters. We
use the same procedure for the $p_{1/2}$ and $d_{5/2}$ states.
\begin{table}
\caption{Energies and widths calculated for low-lying levels of
$^{11}\rm{Be}$ by $S$-matrix pole method. \label{table2}}
\begin{ruledtabular}
\begin{tabular}{ccccccc}
$J^\pi$ & $r_0$ & $a$ & $V_0$  &$V_{ls}$ &$E_{sp}$ & $\Gamma_{sp}$
\\
 &  (fm)&  (fm)&  (MeV) &(MeV)& (MeV)&  (MeV)\\
\cline{1-7} $\frac{1}{2}^{+}$& 1.20 & 0.753 & 57.057 & 0
&-0.503&bound\\
& 1.22 & 0.713 & 57.057 & 0 &-0.503&bound\\
& 1.25 & 0.650 & 57.057 & 0 &-0.503&bound\\
& 1.27 & 0.607 & 57.057 & 0 &-0.503&bound\\
& 1.29 & 0.562 & 57.057 & 0 &-0.503&bound\\ &  &  &  & &  &\\
$\frac{1}{2}^{-}$& 1.20 & 0.819 & 37.505 & 6.0 &-0.183&bound\\
& 1.22 & 0.760 & 37.505 & 6.0 &-0.183&bound\\
& 1.25 & 0.650 & 37.505 & 6.0 &-0.183&bound\\
& 1.27 & 0.545 & 37.505 & 6.0 &-0.183
&bound\\
& 1.28 & 0.451 & 37.505 & 6.0 &-0.183&bound\\ &  &  &  & &  &\\
$\frac{5}{2}^{+}$& 1.20 & 0.753 & 57.057 & 7.131 &1.275&0.221\\
& 1.22 & 0.713 & 57.057 & 6.222 &1.275&0.208\\
& 1.25 & 0.650 & 57.057 & 4.743 &1.275&0.189\\
& 1.27 & 0.607 & 57.057 & 3.671 &1.275&0.176\\
& 1.29 & 0.562 & 57.057 & 2.520 &1.275&0.164\\
\end{tabular}
\end{ruledtabular}
\end{table}
As is seen in Table \ref{table2}, the adopted potential well is shallower
for the $p_{1/2}$ state than for the ground state, which reflects the
inversion of the $s$ and $p$ levels. To reproduce the well-known
energy of $\frac{5}{2}^{+}$ resonance ($E_R=1.275$ MeV), we use the
set of the potential parameters determined for the ground state
$\frac{1}{2}^{+}$ of  ${}^{11}{\rm Be}$ with addition of the
spin-orbital potential. The fact that $\frac{5}{2}^{+}$ state has
particle width provides for an additional criterion for the
selection of the potential. As can be seen in Table \ref{table2},
the calculated single-particle widths for this state are larger
than the experimental values of $100\pm 20$ keV \cite{tunl} and
$104\pm 21$ keV \cite{lui90}. Taking into account that
\begin{equation}
\Gamma_{exp}=S\,\Gamma_{sp}
\label{gamma1}
\end{equation}
we can estimate the spectroscopic
factor $S$ for this state. Here $E_{sp}$ and
$\Gamma_{sp}$ stand for $E_0$ and $\Gamma$, correspondingly.
The experimental
and theoretical spectroscopic factors are in the range 0.45-0.8
\cite{lima07}. The spectroscopic values in the interval 0.45-0.61
are obtained by comparing the data in Table \ref{table2} and the
experimental ones. Taking into account the experimental
uncertainties of 20\%, the highest value of the spectroscopic
factor can be $\sim 0.73$. (To decrease the calculated
single-particle width one has to use a sharper potential (smaller
diffuseness), which seems in contradiction with current
experimental data and the theoretical predictions (see
\cite{gol04} and references therein)). Smaller
experimental uncertainties in the width of the $\frac{5}{2}^{+}$
state result in stronger restrictions in the potential
parameters.

The $S$-matrix pole calculations for the three states, which are all
resonances, for the mirror $^{11}\rm{N}$ nucleus are made using the
potential parameters for the ${}^{11}{\rm Be}$ nucleus by adding
the Coulomb potential of the uniformly-charged sphere of the
radius parameter $r_C$ (Eq.(\ref{GrindEQ__2_})). The results are
shown in Tables \ref{table3} and \ref{table4} for two values of the
radius of the uniformly-charged sphere.
\begin{table}
\caption{Energies and widths calculated for low-lying levels of
$^{11}\rm{N}$ by  $S$-matrix pole method. The Coulomb radius
$r_C=1.1$\,fm. \label{table3}}
\begin{ruledtabular}
\begin{tabular}{ccccccc}
$J^\pi$ & $r_0$ & $a$ & $V_0$  &$V_{ls}$ &$E_{sp}$ & $\Gamma_{sp}$
\\
 &  (fm)&  (fm)&  (MeV) &(MeV)& (MeV)&  (MeV)\\
\cline{1-7} $\frac{1}{2}^{+}$& 1.20 & 0.753 & 57.057 & 0
&1.011&0.832\\
& 1.22 & 0.713 & 57.057 & 0 &1.036&0.869\\
& 1.25 & 0.650 & 57.057 & 0 &1.077&0.931\\
& 1.27 & 0.607 & 57.057 & 0 &1.108&0.978\\
& 1.29 & 0.562 & 57.057 & 0 &1.142&1.032\\ &  &  &  & &  &\\
$\frac{1}{2}^{-}$& 1.20 & 0.819 & 37.505 & 6.0 &1.912&0.936\\
& 1.22 & 0.760 & 37.505 & 6.0 &1.984&0.956\\
& 1.25 & 0.650 & 37.505 & 6.0 &2.126&0.985\\
& 1.27 & 0.545 & 37.505 & 6.0 &2.274
&1.014\\
& 1.28 & 0.451 & 37.505 & 6.0 &2.415&1.035\\ &  &  &  & &  &\\
$\frac{5}{2}^{+}$& 1.20 & 0.753 & 57.057 & 7.131 &3.653&0.946\\
& 1.22 & 0.713 & 57.057 & 6.222 &3.699&0.913\\
& 1.25 & 0.650 & 57.057 & 4.743 &3.772&0.865\\
& 1.27 & 0.607 & 57.057 & 3.671 &3.823&0.834\\
& 1.29 & 0.562 & 57.057 & 2.520 &3.877&0.804\\
\end{tabular}
\end{ruledtabular}
\end{table}

\begin{table}
\caption{The same as in Table \ref{table3} but for the Coulomb radius
$r_C=1.2$\,fm. \label{table4}}
\begin{ruledtabular}
\begin{tabular}{ccccccc}
$J^\pi$ & $r_0$ & $a$ & $V_0$  &$V_{ls}$ &$E_{sp}$ & $\Gamma_{sp}$
\\
 &  (fm)&  (fm)&  (MeV) &(MeV)& (MeV)&  (MeV)\\
\cline{1-7} $\frac{1}{2}^{+}$& 1.20 & 0.753 & 57.057 & 0
&0.997&0.792\\
& 1.22 & 0.713 & 57.057 & 0 &1.022&0.826\\
& 1.25 & 0.650 & 57.057 & 0 &1.062 &0.884\\
& 1.27 & 0.607 & 57.057 & 0 &1.092&0.928\\
& 1.29 & 0.562 & 57.057 & 0 &1.125&0.977\\ &  &  &  & &  &\\
$\frac{1}{2}^{-}$& 1.20 & 0.819 & 37.505 & 6.0 &1.896 &0.912\\
& 1.22 & 0.760 & 37.505 & 6.0 &1.965&0.927\\
& 1.25 & 0.650 & 37.505 & 6.0 &2.102&0.953\\
& 1.27 & 0.545 & 37.505 & 6.0 &2.243
&0.975\\
& 1.28 & 0.451 & 37.505 & 6.0 &2.375&0.990\\ &  &  &  & &  &\\
$\frac{5}{2}^{+}$& 1.20 & 0.753 & 57.057 & 7.131 &3.637&0.932\\
& 1.22 & 0.713 & 57.057 & 6.222 &3.681&0.899\\
& 1.25 & 0.650 & 57.057 & 4.743 &3.752&0.851\\
& 1.27 & 0.607 & 57.057 & 3.671 &3.801&0.819\\
& 1.29 & 0.562 & 57.057 & 2.520 &3.852&0.788\\
\end{tabular}
\end{ruledtabular}
\end{table}
It is worth noting
that in the case of the relatively sharp $\frac{5}{2}^{+}$
resonance, the differences between calculations of the resonance
energy and the width using the $S$-matrix pole and the phase shift
are relatively moderate, 140 keV and 130 keV, correspondingly.
However, these differences become significantly larger for the
broad resonance $\frac{1}{2}^{-}$ in $^{11}\rm{N}$, up to $\sim
300 $ keV for the energy and $\sim 500$ keV for the width. We note
that for the same set of the potential parameters the $S$-matrix pole
method gives energy and width smaller than those obtained by the phase shift.
As for the $2s_{1/2}$ state in $^{11}\rm{N}$, the phase shift never passes
through $\pi/2$ in agreement with the earlier observation by Barker \cite{barker96}.

The $S$-matrix pole method reveals the resonance pole for the state
$\frac{1}{2}^{+}$ in $^{11}\rm{N}$. To check that we have found
the pole correctly, we match the logarithmic derivatives of the
solution of the Schr\"odinger equation and the Gamow function in
the asymptotic region. We also check the ratio of the solution to
the Gamow function, which must be constant at the asymptotic
region. Let us consider the $\frac{5}{2}^{+}$ level in
$^{11}\rm{N}$. By averaging the experimental data from Refs.
\cite{axel96,gold00,guim03,lep98}, we obtain $3.72\pm 0.050$ MeV for the resonance
energy for this level  and $0.55^{+0.05}_{-0.1}$ MeV for the width.
We can conclude from Tables \ref{table3} and \ref{table4}
that the $S$-matrix pole method gives for the width
$\sim 0.85$ MeV resulting in the spectroscopic factor $S= 0.65$.
The standard geometrical parameters $r_0 = 1.25$ fm and $a = 0.65$
fm provide for a good agreement with the average experimental
energy for this level. Using these parameters (Table \ref{table2}),
one can find the spectroscopic factor of 0.53 for the mirror state
in $^{11}\rm{Be}$. Assuming that the spectroscopic factors should
be the same for the mirror states, one can conclude that the
average value of the spectroscopic factor $S=0.59$ is a characteristics of the
single-particle structure for the $\frac{5}{2}^{+}$ state in the
mirror $^{11}\rm{N}$ and $^{11}\rm{Be}$ nuclei.

The experimental data for the broad $\frac{1}{2}^{-}$ resonance
state in $^{11}\rm{N}$ need careful consideration because the results
reported in \cite{axel96,cas06,gold00,guim03} are different
due to different definitions of "energy" and "width" in these works.
We nevertheless conclude that the resonance energy of this
state is $\sim 2.2$ MeV. As in \cite{millenerprivate} we use Eq.
(\ref{gamma1}) to get the width of the $\frac{1}{2}^{-}$ state
in the potential approach. The spectroscopic factor $S$=0.66
results in the width $\Gamma=0.65$ MeV for this state.
This spectroscopic factor coincides with the shell model
prediction for the analog state of $^{11}\rm{Be}$ and our
result is in a good agreement with the one obtained in \cite{millenerprivate}.

All available experimental data \cite{axel96,cas06,gold00} give
higher resonance energies of the $\frac{1}{2}^{+}$ state in
$^{11}\rm{N}$ than our calculations (see Table \ref{table3} and
\ref{table4}). (We exclude most of the mass-transfer data from the
consideration because of the very low population of the
$\frac{1}{2}^{+}$ state in $^{11}\rm{N}$ in these reactions.) The
$^{11}\rm{N}$ ground state resonance energies (the relative
$^{10}\rm{C}+\rm{p}$ energy) are grouped around 1.3 MeV from the
data \cite{cas06,gold00}. The most recent study \cite{gold00}
resulted in the value of 1.54 MeV for the resonance energy; the
experimental widths for the resonance are in the range from 0.83
MeV \cite{gold00} to 1.4 MeV \cite{cas06}. These experimental
values were extracted using different approaches. In
\cite{axel96,cas06} the behavior of the single-particle wave
function inside the $^{11}\rm{N}$ nucleus is used to determine the
resonance energy (identified as the energy, at which the modulus
of the wave function calculated at $1$ fm reaches maximum -
$|\Psi_{max}|$ method) and the resonance width. The $R$-matrix
analysis was used in \cite{gold00}. Both approaches can not
eliminate a contribution from the non-resonant potential
scattering. Leaving a detailed analysis of the experimental data
for future studies, we make a crude estimation of the
spectroscopic factor for the $2s$ state. To this end we average
data from \cite{axel96,cas06} for the $\frac{1}{2}^{+}$  resonance
getting the resonance energy $1.30$ MeV and resonance width 1.20
MeV. The analysis \cite{axel96,cas06} was based on a potential
approach, i.e. the Woods-Saxon potential was found, which allows
fitting the excitation functions and angular distributions for the
elastic proton resonance scattering. Using the potential
parameters from \cite{axel96}, we apply here the $S$-matrix pole
method, rather than the $|\Psi_{max}|$ method, to determine the
resonance energy and width. We obtain $1.102$ MeV for the
resonance energy and $840$ keV for the resonance width, i.e. the
resonance energy of $\frac{1}{2}^{+}$ state in $^{11}\rm{N}$
decreases by $200$ keV compared to the one adopted previously! We
now adopt $1.102$ MeV as a new "experimental" resonance energy of
the $\frac{1}{2}^{+}$ state in $^{11}\rm{N}$.

We note that the potential found in \cite{axel96} does not reproduce the experimental binding energy of the $\frac{1}{2}^{+}$ state and resonance energy of $\frac{5}{2}^{+}$ state in ${}^{11}{\rm Be}$. Meanwhile the potentials given in Tables \ref{table3} and \ref{table4} fit $2s$ and $1d$ states both in $^{11}\rm{Be}$ and $^{11}\rm{N}$. Then, we assume that the potential with the standard geometry $r_0=1.25$ fm and $a=0.65$ fm in Table \ref{table4} is the "right" one. Note that the resonance energy and width obtained for this potential are very close to the average resonance energy and width shown in Table \ref{table4} for 5 different potentials. This potential gives $1.062$ MeV  resonance energy, which is a pure single-particle energy. We observe that this energy is $\sim 0.04 $ MeV less than the "experimental" value of $1.102$ MeV obtained for the potential adopted in \cite{axel96}. This $0.04$ MeV can be attributed to the non-single-particle admixture to the structure of the $\frac{1}{2}^{+}$ state $^{11}\rm{N}$.

An estimation of the spectroscopic factor can be obtained
from the consideration of the width of the state. The $r_0=1.25$ fm and $a=0.65$ fm
parameters generate 0.95 MeV for the $\frac{1}{2}^{+}$ state width at the "experimental"
resonance energy of 1.10 MeV. The ratio of 0.84/ 0.95
(the "experimental" width/ calculated width) results in the spectroscopic
factor of 0.88 for the adopted potential with the standard geometry.
Hence we obtain much larger spectroscopic factors for the ground state in $^{11}\rm{N}$
than for the $\frac{5}{2}^{+}$ excited state.

As a final remark to this section, it is worth noting that the
conventional potential approaches, which determine the resonance
energy and width from the energy dependence of the phase shift or
from the $|\Psi_{max}|$ method, may not give accurate results
because of the distortion generated by the non-resonant background
at physical energies. For example, $|\Psi_{max}|$ may reach a peak
in the internal region at energy $E \not= E_{0}$. In this sense
the $S$-matrix pole is the most accurate method for a given
potential because it determines the resonance energy and width by
searching the resonant pole at complex energy, i.e. separates the
resonant contribution from the background. We note that the
resonance energy determined by the $S$-matrix pole method depends on
the adopted potential. Moreover, the resonance parameters
determined by the $S$-matrix pole method may differ from the ones
determined from the $R$-matrix approach as we have seen it for the ${}^{15}{\rm F}$
case.

\subsubsection{Model-independent determination of the energy and width of the broad
resonance $\frac{1}{2}^{+}$ in $^{11}\rm{N}$}

Here we determine the resonance parameters for the $2s_{1/2}$
resonance in ${}^{11}{\rm N}={}^{10}{\rm C} +p$ using the
model-independent representation of the elastic scattering
$S$-matrix given by Eq. (\ref{elsmatr1}). We use the Woods-Saxon
potential with the standard geometry $r_{0}=1.25$ fm, $a=0.65$ fm,
the Coulomb radial parameter $r_{C}=1.2$ fm and the depth $V_{0} =
57.06$ MeV to generate the "quasi-experimental" $2s_{1/2}$ phase
shift. The resonance energy for this  potential obtained from the
Schr\"odinger equation is $E_{0}= 1.062$ MeV and the resonance
width $\Gamma= 0.884$ MeV  (see Table \ref{table4}). To fit this
"quasi-experimental" phase shift  we use the polynomial approximation  
$\delta_{p}(k)= \sum\limits_{n = 0}^3 {{b_n}{{(k - {k_s})}^n}}$ in 
Eq. (\ref{elsmatr1}).  The $S$-matrix pole method based on Eq. (\ref{elsmatr1}) 
gives $E_{0}=1.057$ MeV and $\Gamma=0.880$ which agrees extremely well
 with the potential $S$-matrix pole. As starting search value in fitting the
"quasi-experimental" phase we used $k_{s}= 0.25$ fm$^{-1}$ but the
result only depends slightly on the initial $k_{s}$ value.

\subsection{The subthreshold resonances in the proton--proton system}

The poles for the antibound state of the singlet
neutron-neutron or the neutron$-$proton systems are located on the
imaginary axis in the complex momentum plane (at energies
$E_{nn}\cong -134 $ keV and $E_{np}\cong -66$ keV ).
In  \cite{kok80} using the effective-range approach  Kok showed that
in the case of the proton-proton system the pole moves to the complex plane,
due to the Coulomb barrier. The ground state pole of the $s$-wave \textit{ pp}
scattering amplitude was found in \cite{kok80} at $k_{pp} =(0.0647 -i
0.0870)$ fm$^{-1}$ or $E_{pp}=(-140-i 467)$ keV. The effective-range
parameters for the standard expansion were taken from
\cite{bro76}. Recently, calculations with the same
approximation were repeated in \cite{yer07} resulting in $k_{pp} =
(0.0644 - i 0.0871)$ fm$^{-1}$ or $E_{pp}=(-142-i 465)$ keV, which
is in a good agreement with the Kok's result.

A definition of the renormalized partial amplitude in the presence
of the Coulomb interaction was given earlier (see Eq. (3) in \cite{blok84}).
A new corresponding formula was derived in \cite{yer07} for the renormalized vertex constant $G_{ren}$ for the
virtual decay of a nucleus into two charged particles in the effective-range theory. It was applied to the \textit{pp}
and \textit{pd} systems using the standard effective--range
expansion and the effective--range function with a pole,
respectively. The value of $G_{ren}^2$ is real quantity for the
bound state because the energy is real. In the case of
the resonance, the energy is complex so $G_{ren}^2$ becomes complex.
For the \textit{ pp } ground state, the value $G_{ren}^2 =(0.060+i
0.051)$ fm was obtained in \cite{yer07}  with the effective--range parameters
taken from \cite{bro76}. The only condition which validates these
results is the convergence of the effective-range expansion near
the pole considered. It was shown in \cite{kok80} that
the results change only slightly when the parameters of form are
neglected (\textit{P}=\textit{Q}=0). The convergence is ensured
in the case of the \textit{ pp } subthreshold resonance pole.

Nevertheless, the effective-range approximation has some
drawbacks. It gives the partial scattering amplitude in an
analytical form as a ratio of two polynomials. As a result, all
the amplitude singularities are poles in the complex momentum
plane. The number \textit{n} of the poles is obviously defined by
the maximal degree used in the effective range expansion up to
$k^n$, which gives the degree of Kok's equation for the position of
the pole and correspondingly the number of its solutions without the Coulomb
force. For example, a logarithmic dynamical
cut of the amplitude in the case of the two-body model with the Yukawa
potential can not be reproduced in this approximation. But it is
imitated by a pole located on the positive imaginary axis, which is not a bound
state pole. In the case of the \textit{pp} system, the situation is simple 
because there is no bound state, so any pole on the positive energy axis is an
unphysical one. Moreover, the region of the validity of the
effective-range approximation is limited by the condition $|k|\leq
|k_{max}|$, where the effective-range expansion converges. In the 
potential model with the asymptotic $V(r)\rightarrow const\cdot
r^\nu\cdot exp\left(-r/R\right)$, the value $|k_{max}|=1/(2R)$ is
the beginning of the dynamical cut on the imaginary axes in the
complex \textit{ k } plane. In the case of charged particles, the
number of roots is infinite (see \cite{orl06}). In particular, as
noted in \cite{orl06}, the sequence of poles located near the
negative imaginary axis can be mistakenly identified as virtual 
(antibound) state poles known for the system without the Coulomb 
interaction.

Finding the pole by solving the Schr\"odinger equation is the most
reliable way to confirm that the pole found by a solution of Kok's
equation \cite{kok80} is not a false one. This was done in the
present paper for the \textit{pp} system with the Yukawa
potential. Its parameters are taken from \cite{bro76} for the
singlet \textit{np} system. We find that $k_{pp}^{Yu} = (0.064
-i\,0.082) \rm{fm}^{-1}$ (or $E_{pp}=(-106.7-i\, 435.5)$ keV ) for
the \textit{pp} ground state. After that, we have slightly changed
the geometric parameter to describe the experimental $pp$
scattering length and effective range and the resolved
Schr\"odinger equation gives $E_{pp}=(-138.16-i\, 463.14)$ keV.
This result almost coincides with Kok's results. The pole for $nn$
system is $E_{nn}=-92$ keV. The resonance wave function contains
the outgoing wave in the asymptotic region while the ingoing wave
is absent. For the normalization of the wave function in this case
we can not use the Zeldovich procedure because
$\rm{Re}(k_{pp})<\rm{Im}(k_{pp})$, however, we can find residue at
the pole. The residue in the pole corresponding to the
subthreshold resonance is $A_{pp}=(-0.021+ i0.057)\,\,
\rm{fm}^{-1}$.

\section{CONCLUSION}

In the present work we apply the $S$-matrix pole method to determine
the energies of the bound, the virtual states and
resonances. This method is based on a numerical solution of the
Schr\"odinger equation. Usually this method is applied to the bound
states, but here it is extended to the resonance and virtual states
despite the fact that the corresponding solutions increase exponentially
when $r\rightarrow\infty$. The method turns out to be especially
useful for broad resonances including subthreshold ones.

There can be a few  poles in the complex plane when applying the effective 
range theory and the  number of the poles is defined by the maximal 
power $k^{n}$ used in the effective range expansion.  An additional investigation 
should be done to select a physical pole. In our approach one gets no false 
poles, thus resolving ambiguity problem appearing in the effective range approach.

In the case of the resonances potential models and $R$-matrix
approach are commonly used to analyze the experimental data, and
resonance parameters are determined from the fits. For narrow
resonances both approaches give accurate results. However, this is
not the case for broad resonances. In this case, due to the
distortion caused by the non-resonant background at physical
energies, the resonance energy and the width determined from the
fitting of the experimental data depend on the model and within a given
model the prescriptions to determine the resonance energy and width may be 
different. Usually researchers use different definitions of the resonance 
energy and width. Broad resonance parameters extracted from the experimental 
data are model dependent. For this reason, one should indicate the method 
used to determine them in any subsequent references.

Here we address two methods for determining the
resonance energy and width from the pole of the $S$-matrix: the
potential $S$-matrix pole method based on the solution of the Schr\"odinger
equation and the $S$-matrix pole method based on the analytical continuation for the $S$-matrix to the resonant pole. We compare the results for the resonance parameters obtained from the different determinations of the resonance energy and width in the potential approach, the $S$-matrix pole methods and $R$-matrix method. Correct evaluations of the resonance parameters are important when comparing the experimental data, both for the tests  of the isobaric multiplet mass equation and for detailed structure calculations of the exotic nuclei. The potential $S$-matrix pole method provides the most accurate resonance energy and width for a given potential. The second $S$-matrix pole method, which uses Eq. \ref{elsmatr1}, is even more general because it does not require any potential model and is based only on the analyticity and symmetry of the $S$-matrix. In contrast to other approaches, the pole $S$-matrix methods allow one to correctly separate the resonance pole contribution and the nonresonant background.  

Our approach has a potential of being extended to treat broad resonance populated in transfer reactions, where the half-off-energy shell resonant amplitude interferes with the half-off-energy shell nonresonant amplitude. At present there is a huge disagreement in the resonance parameters for broad resonances obtained from the resonance or direct reactions \cite{cas06}.

\section*{Acknowledgments}
The work was supported by the U.S. Department of Energy under Grant No. DE-FG02-93ER40773 and DE-FG52-06NA26207, NSF under Grant No. PHY-0852653, the Russian Foundation for Basic Research, project No. 07-02-00609 and the HEC of Pakistan under Grant No. 20-1171-R.

\end{document}